\documentclass{webofc}
\usepackage[varg]{txfonts}   
\usepackage{hyperref}
\usepackage{graphicx}
\usepackage{subcaption}

\usepackage{color}
\usepackage[capitalise,noabbrev]{cleveref}
\usepackage[newfloat=true,frozencache,cachedir=minted-cache]{minted}
\setminted{fontsize=\scriptsize,linenos}
\usepackage[most,minted]{tcolorbox}
\setminted{fontsize=\footnotesize}

\usepackage{lineno}

\usepackage{tikz}
\usetikzlibrary[decorations]
\usetikzlibrary{arrows}
\usetikzlibrary{calc}
\usetikzlibrary{decorations.markings}
\usetikzlibrary{decorations.text}
\usetikzlibrary{decorations.pathmorphing}
\usetikzlibrary{decorations.pathreplacing}
\usetikzlibrary{shapes}
\usetikzlibrary{shadows}
\usetikzlibrary{positioning}
\usetikzlibrary{mindmap,trees}	
\usetikzlibrary{overlay-beamer-styles}  

\tikzset{%
  far/.style={postaction=decorate,decoration={markings, mark=at position 0.5 with {\arrow[]{>}}}},
  fbar/.style={postaction=decorate,decoration={markings, mark=at position 0.5 with {\arrow[]{<}}}},
  shadowed/.style={postaction={draw=white,color=white,very thick,fill=white}},
  block/.style={
    draw,
    rectangle,
    minimum height=1.5cm,
    minimum width=3cm, align=center
  },
  line/.style={->,>=latex'},
  linew/.style={->,>=latex',line width=0.8pt},
  layer/.style={rounded corners,draw=#1!50!black!50,fill=#1!80!black!10,align=center},
  layern/.style={draw=#1!30!black!30,fill=#1!30!black!10,align=center}
}

\usetikzlibrary{fit}
\tikzstyle{decision} = [diamond, draw, text badly centered, node distance=2.8cm]
\tikzstyle{block} = [rectangle, draw, text centered, rounded corners, text width=1.9cm, node distance=2.2cm]
\tikzstyle{autoblock} = [rectangle, draw, text centered, rounded corners, node distance=2.2cm]
\tikzstyle{line} = [draw, -triangle 90]
\tikzstyle{dline} = [draw, dashed, -triangle 90]

\newcommand{\linkk}[2]{\href{#1}{#2}\xspace}

\begin{document}
\title{Key4hep: Progress Report on Integrations}
%
%

\author{\firstname{Erica} \lastname{Brondolin}\inst{1}
  \and \firstname{Juan Miguel} \lastname{Carceller}\inst{1}
  \and \firstname{Wouter} \lastname{Deconinck}\inst{2}
  \and \firstname{Wenxing} \lastname{Fang}\inst{3}
  \and \firstname{Brieuc} \lastname{Francois}\inst{1}
  \and \firstname{Frank-Dieter} \lastname{Gaede}\inst{4}
  \and \firstname{Gerardo} \lastname{Ganis}\inst{1}
  \and \firstname{Benedikt} \lastname{Hegner}\inst{1}
  \and \firstname{Clement} \lastname{Helsens}\inst{1}\fnsep\inst{5}\fnsep\thanks{Now at EPFL, Lausanne, Switzerland}
  \and \firstname{Xingtao} \lastname{Huang}\inst{6}
  \and \firstname{Sylvester} \lastname{Joosten}\inst{7}
  \and \firstname{Sang Hyun} \lastname{Ko}\inst{8}
  \and \firstname{Tao} \lastname{Lin}\inst{3}
  \and \firstname{Teng} \lastname{Li}\inst{6}
  \and \firstname{Weidong} \lastname{Li}\inst{3}
  \and \firstname{Thomas} \lastname{Madlener}\inst{4}
  \and \firstname{Leonhard} \lastname{Reichenbach}\inst{1}\fnsep\inst{9}
  \and \firstname{Andr\'e} \lastname{Sailer}\inst{1}\thanks{\email{andre.philippe.sailer@cern.ch}}
  \and \firstname{Swathi} \lastname{Sasikumar}\inst{1}
  \and \firstname{Juraj} \lastname{Smiesko}\inst{1}
  \and \firstname{Graeme A} \lastname{Stewart}\inst{1}
  \and \firstname{Alvaro} \lastname{Tolosa-Delgado}\inst{1}
  \and \firstname{Valentin} \lastname{Volkl}\inst{1}
  \and \firstname{Xiaomei} \lastname{Zhang}\inst{3}
  \and \firstname{Jiaheng} \lastname{Zou}\inst{3}
}
\institute{
  CERN, Geneva, Switzerland
  \and University of Manitoba, Winnipeg, Manitoba, Canada
  \and IHEP Beijing, China
  \and Deutsches Elektronen-Synchrotron DESY, Germany
  \and Karlsruhe Institute of Technology, Karlsruhe, Germany
  \and Shandong University, China
  \and Argonne National Laboratory, Lemont, Illinois, USA
  \and Seoul National University, Korea
  \and University of Bonn, Germany
}

\abstract{%
Detector studies for future experiments rely on advanced software tools to estimate performance and optimize their
design and technology choices. The Key4hep project provides a flexible turnkey solution for the full experiment
life-cycle based on established community tools such as ROOT, Geant4, DD4hep, Gaudi, podio and spack. Members of the
CEPC, CLIC, EIC, FCC, and ILC communities have joined to develop this framework and have merged, or are in the progress
of merging, their respective software environments into the Key4hep stack.

These proceedings will give an overview over the recent progress in the Key4hep project: covering the developments
towards adaptation of state-of-the-art tools for simulation (DD4hep, Gaussino), track and calorimeter reconstruction
(ACTS, CLUE), particle flow (PandoraPFA), analysis via RDataFrame, and visualization with Phoenix, as well as tools for
testing and validation.
}
\maketitle
\section{Introduction}\label{sec:intro}

Detector studies for future experiments require advanced software tools to optimize their
design and technology choices and to estimate their performance. These advanced software tools must include the possibility for full or parameterized detector
simulation, the reconstruction of tracks and calorimeter clusters, jet clustering, flavour tagging, and analysis. A
combined solution for all these issues should also allow experiments to move seamlessly from different stages of their
life cycle: for example, from parameterized detector studies to find a performance envelope of their experiment to full
and detailed simulation and reconstruction studies to confirm the validity and feasibility of the assumptions used in
the parameterized simulation. The consistency of the framework also allows experiments to extract performance
parameterizations from detailed simulation studies to create large scale
samples that are not feasible for small communities with limited computing resources available to them.

The Key4hep project provides a solution for these use-cases by means of a structured software stack, which integrates
individual packages towards a complete data processing framework for HEP experiments. The sharing of common components
will reduce the overhead faced by different communities otherwise. Moreover, Key4hep aims to provide an easy-to-use product for
librarians, who provide software installations, the developers creating or adapting components, and the endusers.

\begin{figure}[tbp]
  \centering
  \includegraphics[width=0.5\textwidth]{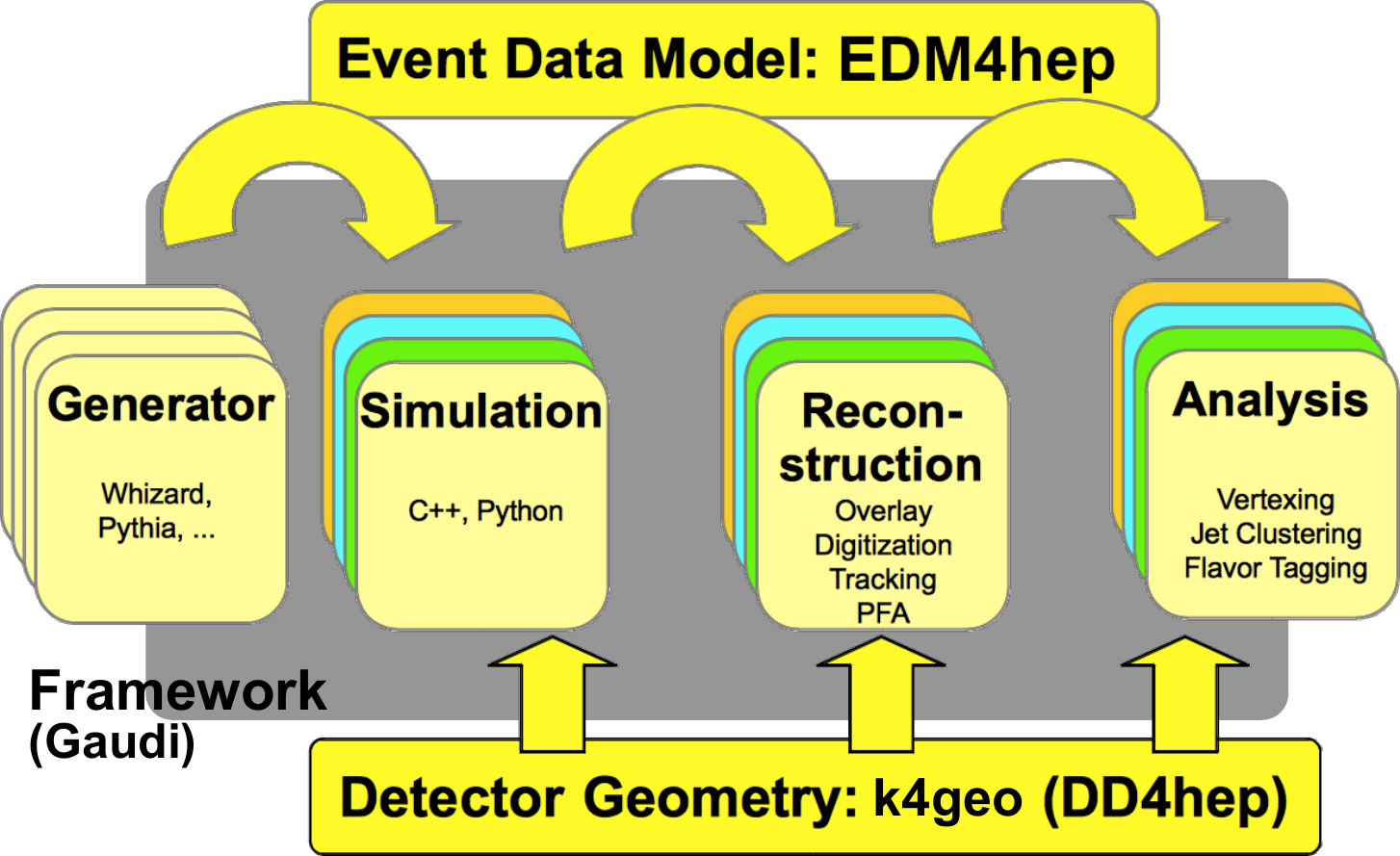}
  \caption{Main ingredients for the Key4hep project: geometry information, event data information, and a processing
    framework with a large number of algorithms.\label{fig:dataflow}}
\end{figure}

\Cref{fig:dataflow} schematically shows the three major ingredients for the project: a
processing framework that connects all the pieces; a way to describe the geometry of the experiments and use the
information for simulation, reconstruction or analysis; and an event data model to exchange data between the pieces,
or for persistency. For Key4hep
the processing framework is Gaudi~\cite{Barrand:2001ny}, the geometry information is provided via
DD4hep~\cite{zenodo_dd4hep_all,dd4hep}, and the event data model is provided by
EDM4hep~\cite{zenodo:edm4hep_all,Gaede:2021izq,Gaede:2022leb} and podio~\cite{podio,podio_2020}.

The current contributors and users of the Key4hep ingredients are part of the CEPC, CLIC,
EIC~\cite{chep23_eic_software}, ILC, FCC, and Muon collider communities. The source code for all components under direct
development of the Key4hep project is hosted on
GitHub\footnote{\href{https://github.com/key4hep}{https://github.com/key4hep}}. Weekly open meetings are held to
discuss ongoing developments and issues in Key4hep. Newcomers are always welcome to join the meetings or contribute to
the developments.

As the goal of the project is not to develop all components itself, but as much as possible reuse existing
solutions, this paper, in \cref{sec:integration}, describes the current status of the integration of tools for simulation,
reconstruction and analysis, then shows, in \cref{sec:testing}, how some of the testing for the project is done, before
it ends with a summary and outlook in \cref{sec:summary}.

\section{Integrations}\label{sec:integration}

\begin{figure}[tbp]
  \centering
  \tikzset{
block/.style={
    draw, 
    rectangle, 
    minimum height=1.5cm, 
    minimum width=3cm, align=center
    }, 
    line/.style={->,>=latex'},
    layer/.style={rounded corners,draw=#1!50!black!50,fill=#1!80!black!10,align=center},
    layern/.style={draw=#1!30!black!30,fill=#1!30!black!10,align=center}
}
\begin{tikzpicture}[scale=0.9,transform shape]
\node[block, layer=white] (a1) {Prev.\\ algorithm\hspace*{20pt}};
\node[block, layern=green, rotate=-90, minimum height=1cm, minimum width=1.5cm, yshift=1.25cm] (a1out) {\footnotesize EDM4hep\\\footnotesize Output};

\node[block, layer=white, right=1cm of a1] (a2) {\hspace*{30pt}MarlinProcessorWrapper\hspace*{30pt}};
\node[block, layern=green, rotate=-90, minimum height=1cm, minimum width=1.5cm, yshift=3cm] (a2in) {\footnotesize LCIO\\\footnotesize Input};
\node[block, layern=green, rotate=-90, minimum height=1cm, minimum width=1.5cm, yshift=8.25cm] (a2out) {\footnotesize LCIO\\\footnotesize Output};

\node[block, layer=white, right=1cm of a2] (a3) {\hspace*{20pt}Next alg.\\\hspace*{20pt}e.g. ACTS};
\node[block, layern=green , rotate=-90, minimum height=1cm, minimum width=1.5cm, yshift=10cm] (a3in) {\footnotesize EDM4hep\\\footnotesize Input};

\node[block, layer=red, below=1cm of a1, xshift=2.0cm] (c1) {EDM4hep2LCIO\\ converter};

\node[block, layer=red, below=1cm of a3, xshift=-1.8cm] (c2) {LCIO2EDM4hep\\ converter};

\draw[dashed] (a1out.north) -- (a2in.south);
\draw[dashed] (a2out.north) -- (a3in.south);



\path[->, bend right, out=0, in=270] (a1out.east) edge (c1.west);
\path[->, bend left, out=270, in=180] (c1.east) edge (a2in.east);

\path[->, bend right, out=0, in=270] (a2out.east) edge (c2.west);
\path[->, bend left, out=270, in=180] (c2.east) edge (a3in.east);

\end{tikzpicture}
  \caption{Schematic of how Marlin processors are integrated into Gaudi workflows in Key4hep.\label{fig:inserter}}
\end{figure}
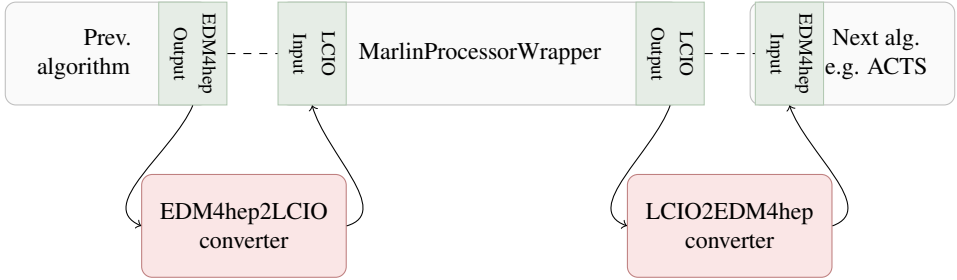

One part of the integrations into Key4hep are the existing experiment software components from CEPCSW and FCCSW, which
are already based on Gaudi. Their adaptation to Key4hep required mostly adaptations to the EDM4hep event data
model~\cite{Volkl:20225e,FernandezDeclara:2022voh,Fang:2023mwt}. For the integration of iLCSoft, used by the ILC and
CLIC communities, the k4MarlinWrapper~\cite{zenodo:k4marlinwrapper_all} was created to integrate \emph{processors} from the Marlin
framework~\cite{Gaede:2006pj} and its corresponding event data model LCIO~\cite{Gaede:2003ip, chep23_lcio} into Gaudi. The idea of the
k4MarlinWrapper is shown in \cref{fig:inserter}. To run any \emph{processor} from Marlin in a Gaudi workflow, the event
data is converted in memory from EDM4hep to LCIO before the execution of the processor and back to EDM4hep after.

In the following sections, the status of the integration for simulation, reconstruction and analysis tools is laid out.

\subsection{Simulation}\label{sec:sim}

For full or parameterized simulation, different tools are already available in Key4hep. The parametric simulation program
Delphes~\cite{deFavereau:2013fsa}, together with some utilities to handle the generation of primary events, has been
integrated into Gaudi as k4SimDelphes~\cite{zenodo:k4simdelphes_all}. k4SimDelphes also contains standalone programs for different input files, such as HepMC, or controlling event
generators, such as Pythia8. There are two possibilities for full detector simulation with Geant4~\cite{Geant4} and the
DD4hep geometry. There is the ddsim~\cite{ddsim} feature of DD4hep, which can produce EDM4hep output files and read a majority of
generator output formats. In addition, there is the k4SimGeant4~\cite{zenodo:k4simgeant4_all} Gaudi integration, which came out
of FCCSW. The framework integration of the full simulation via k4SimGeant4 -- together with the other algorithms --
allow one to run a complete chain from event generation to reconstructed objects in a single program execution.

In the meantime, also the Gaussino~\cite{lhcbdd4hep2019} functionalities from LHCb have become experiment
agnostic~\cite{chep23_gaussino} and will potentially provide a complete replacement of the k4SimGeant4 package.

\subsection{Reconstruction}\label{sec:reco}

\subsubsection{Tracking}\label{sec:rec_track}

\begin{figure}[tbp]
  \centering
  \includegraphics[width=0.50\textwidth]{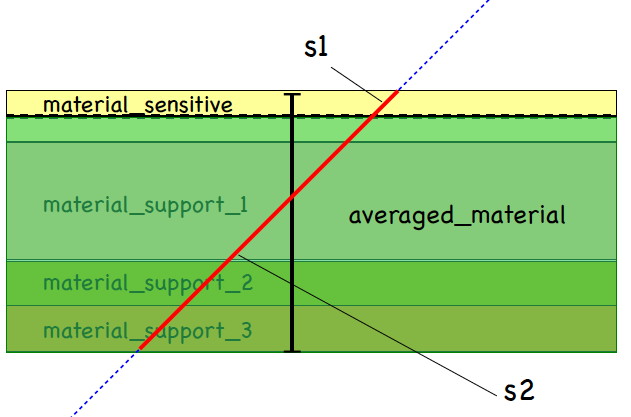}
  \caption{Sketch showing the averaging of materials when creating \texttt{DD4hep::rec::Surfaces}.\label{fig:ddsurface}}
\end{figure}

The iLCSoft ecosystem contains tools for track pattern recognition and track fitting, all of which can be used in Key4hep via the k4MarlinWrapper
For the track fitting \texttt{DD4hep::rec::Surfaces} are attached to any sensitive element~\cite{sailer17:ddrec} and some dead
material. These surfaces provide an abstract view of the geometry needed for reconstruction, such as automatically averaged materials
(as shown in \cref{fig:ddsurface}) and measurement directions. In many cases, these surfaces can be pragmatically
attached to an existing DD4hep based geometry using run-time plugins. This easily enables track reconstruction for different detectors.

The ACTS~\cite{ai22:_common_track_softw_projec,Gessinger-Befurt_2023} integration into Key4hep progresses steadily. In
the last months, a plugin to support the EDM4hep track format was added to ACTS~\cite{chep23_acts}. Support for DD4hep
geometry is under active development as well. It is already possible to load DD4hep geometries following a certain
hierarchical structure into ACTS\@. The existing conversion of DD4hep to ACTS geometry is currently heavily used by the
ACTS developers to test their algorithms against the generic Open Data Detector, a detector model used for benchmarking
tracking and calorimeter reconstruction approaches~\cite{chep2023_odd}. However, the general use of
\texttt{DD4hep::rec::Surfaces} is under development and will allow a broader range of detectors to be directly
used. Using these surfaces also in the link to the ACTS reconstruction will enable a direct replacement of
the iLCSoft track reconstruction with ACTS\@. Recently, the development of the necessary Gaudi algorithms to use ACTS
from Key4hep has intensified. In particular, the focus is on enabling arbitrary track refits using different fitters via
Gaudi.

\subsubsection{Calorimeter Clustering}\label{sec:rec_calo}

An important ingredient for the performance of future Higgs Factory experiments is the particle flow reconstruction for
optimal jet energy resolutions. The Pandora particle flow algorithm package
(PandoraPFA)~\cite{Thomson_2009,Marshall:2015rfaPandoraSDK} was developed to study particle flow clustering at linear
colliders.

The particle flow clustering with Pandora makes use of the extensions attached to detector
geometries~\cite{sailer17:ddrec}, such as \texttt{DD4hep::rec::LayeredCalorimeterData}, to provide the properties of the
calorimeter, e.g., radiation length, interaction length, and dimensions to the reconstruction algorithms. To support a
larger range of detectors, for example those that foresee a noble liquid calorimeter~\cite{Francois:2835881}, the necessary information will be
obtained in a more dynamic way. This step was materialized using the \texttt{DD4hep::MaterialManager} to extract the
necessary information between arbitrary space points.

\begin{figure}[tbp]
  \centering
  \subcaptionbox{Simulated photon clusters shown with CED (C-Event display, see \cref{sec:vis}).\label{fig:sim_clusters}}{\includegraphics[width=0.49\textwidth]{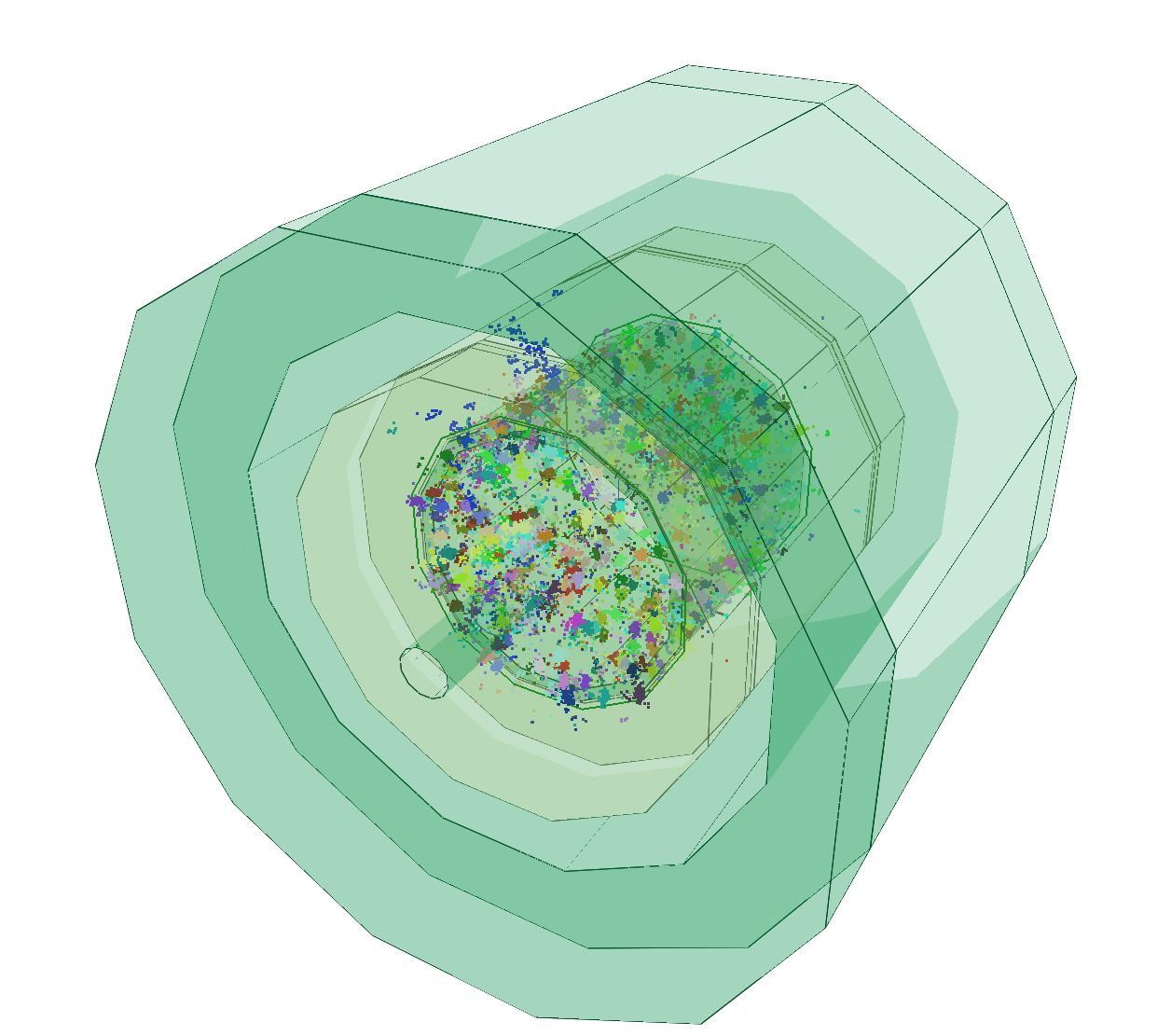}}
  \subcaptionbox{Reconstructed clusters from
    CLUE.\label{fig:rec_clusters}}{\includegraphics[width=0.49\textwidth]{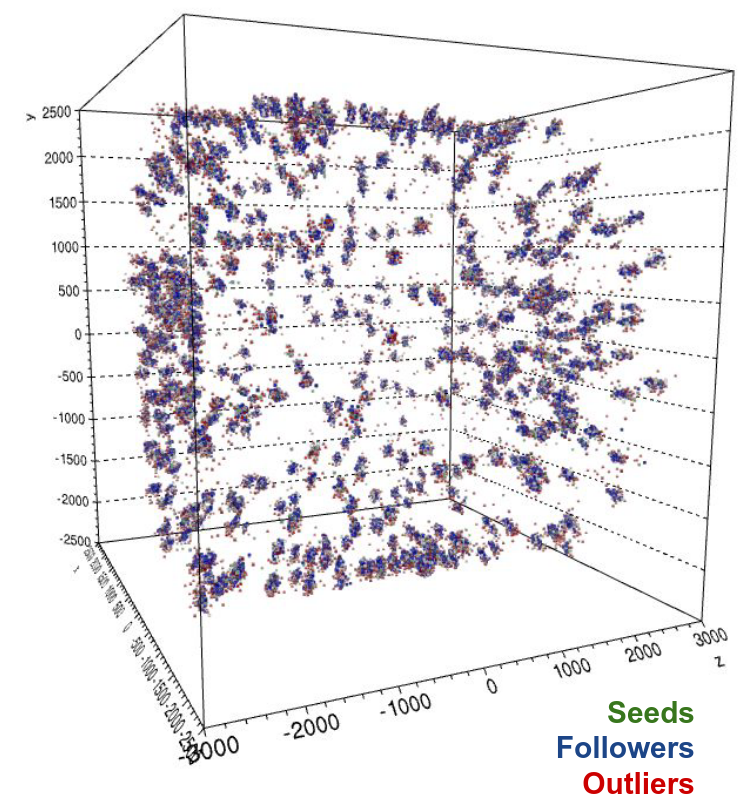}}
  \caption{Simulated photons and their clusters reconstructed with k4clue.\label{fig:k4clue}}
\end{figure}

At least for high granularity calorimeters with large occupancies, the reconstruction time can become
dominant. For the HGCAL project of CMS a GPU friendly algorithm, CLUE (CLUstering of Energy), was developed~\cite{rovere2020clue,Brondolin_2023}.
An integration of this algorithm in Key4hep is ready to be used~\cite{brondolin_erica_2023_8256333}. \Cref{fig:k4clue}
shows simulated hits from many photons in a single event, and how CLUE reconstructs them into clusters.

\subsection{Analysis}\label{sec:analysis}

The ROOT persistency of the EDM4hep event data model and its columnar storage model allows the use of the
RDataFrame~\cite{Piparo:2019xdy} feature for analysis. \Cref{lst:dataframe} shows an example how the EDM4hep objects,
here \texttt{ReconstructedParticles}, can be used with a dataframe to select events with different criteria. To take full
advantage of the EDM4hep datamodel based on podio, for example to handle relationships between different
objects, an \texttt{RDataSource} is being developed.

\begin{listing}[btp] 
\begin{minted}{Python}
theDataFrame
# define an alias for electron index collection
.Alias("Electron_idx", "Electron_objIdx.index")
# define the electron collection
.Define("electrons",  "ReconstructedParticle::get(Electron_idx, ReconstructedParticles)")
#select electrons on pT
.Define("selected_electrons", "ReconstructedParticle::sel_pt(10.)(electrons)")
# ...
.Define("zed_leptonic_recoil_m","ReconstructedParticle::get_mass(zed_leptonic_recoil)")
# create branch with leptonic charge
.Define("zed_leptonic_charge","ReconstructedParticle::get_charge(zed_leptonic)")
# Filter at least one candidate
.Filter("zed_leptonic_recoil_m.size()>0")
\end{minted}
  \caption{Example for the usage of EDM4hep objects for analysis with RDataFrame~\cite{fccanalyses:github}.\label{lst:dataframe}}
\end{listing}

\subsection{Visualization}\label{sec:vis}
\begin{figure}[tbpp]
  \begin{minipage}{0.5\textwidth}
    \includegraphics[width=1.0\textwidth]{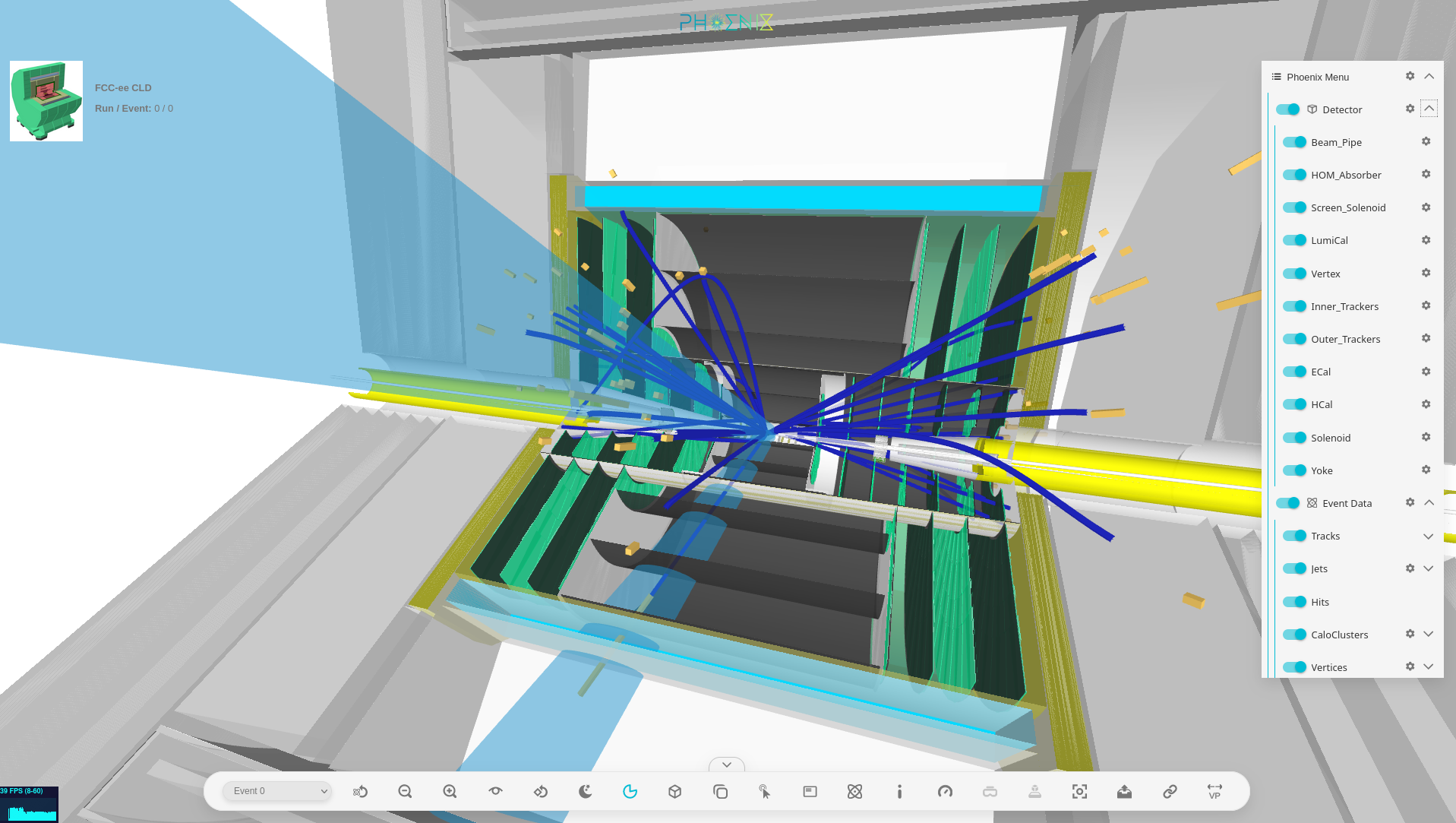}
    \caption{An event display of the CLD detector using Phoenix.\label{fig:phoenix}}
  \end{minipage}
  \begin{minipage}{0.5\textwidth}
    \includegraphics[width=1.0\textwidth]{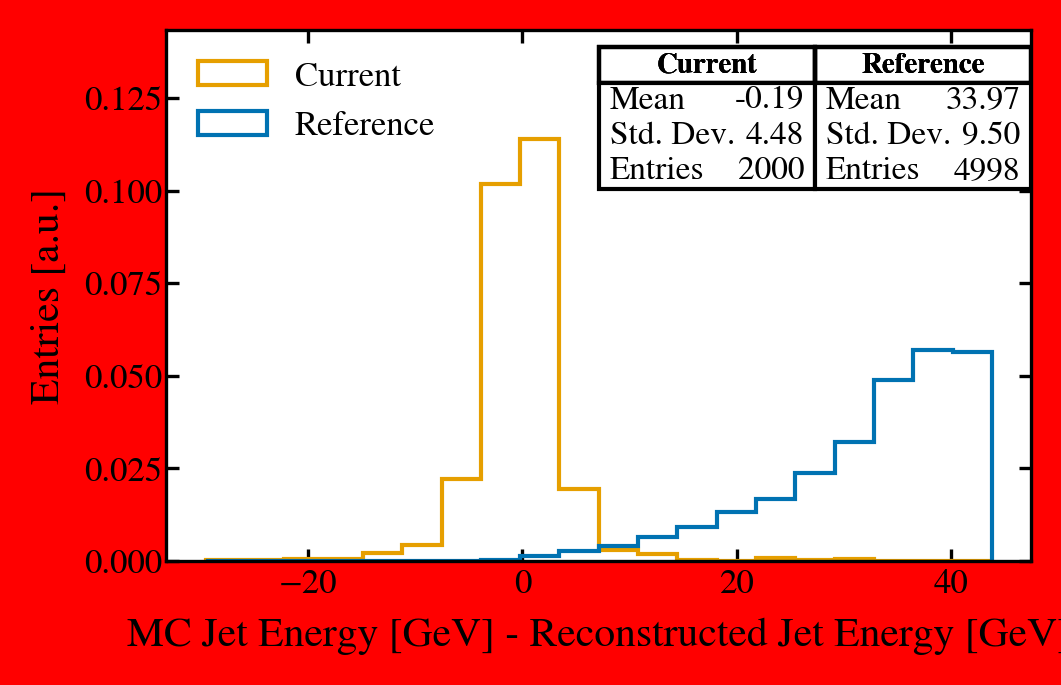}
  \caption{Bright red highlighting differing distributions. \emph{Current} is after a bug was
    fixed.\label{fig:val_fail}}
\end{minipage}
\end{figure}

The DD4hep geometry, and the EDM4hep event data can both be converted to formats suitable for the Phoenix event
display~\cite{zenodo:phoenix_all}. \Cref{fig:phoenix} shows an event in the CLD detector for the FCCee~\cite{CLD}, which is also shown in
\cref{fig:sim_clusters} using the C-Event display (CED) from iLCSoft. The advantage of Phoenix is the possibility of using a
web-browser, and its broad configurability. Phoenix allows one to centrally host the detectors on the web. The FCC
detectors, for example, are hosted on a webserver\footnote{\url{https://fccsw.web.cern.ch/fccsw/phoenix/}} and continuously updated.

\section{Testing}\label{sec:testing}

A continuous validation system has been set up for Key4hep. Every night, after the nightly build of the Key4hep stack is
finished, detector simulation is done based on the preset detectors in the configuration. After that, a complete
reconstruction is performed. The results of the reconstruction are then compared to a set of reference samples that were
produced in known and reproducible conditions. If the new distributions are different from the reference ones, based on
specified metrics, then the display of the plots in a webpage will point this out (see \cref{fig:val_fail}). The plots in the webpage are
classified in different categories depending on which class of results they belong to, for example plots about tracks in
one category, those about jet reconstruction in another. The current system is evolving and several
improvements are under development, such as easier configuration, more detectors being tested and a better reporting
system when the new and reference distributions are different.

Besides the physics performance, the CPU performance has to be monitored and controlled as well. For this purpose
the \emph{Valprod} toolkit is under development, which enables the building of comprehensive validation jobs and offers CPU
flame graph, I/O profiling and an integration with the
prmon~\cite{zenodo:prmon_all} program.

\section{Summary \& Outlook}\label{sec:summary}

\begin{figure}[tbp]
  \centering
  \begin{tikzpicture}[scale=0.7,transform shape]
  \Large
  \node[draw, rectangle, minimum height=6cm, minimum width=4cm, align=center, layer=white, xshift=0cm] (a1) {};
  \node[draw, rectangle, minimum height=6cm, minimum width=8cm, align=center, layer=white, xshift=7cm] (a2) {};
  \node[draw, rectangle, minimum height=6cm, minimum width=4cm, align=center, layer=white, xshift=14cm] (a3) {};

  \node[draw, rectangle, minimum height=2cm, minimum width=3cm, align=center, layer=green, yshift= 1.5cm] (a1s1) {MCGenerator\\Output Files};
  \node[draw, rectangle, minimum height=2cm, minimum width=3cm, align=center, layer=green, yshift=-0.0cm] (a1s2) {Particle Gun};
  \node[draw, rectangle, minimum height=2cm, minimum width=3cm, align=center, layer=green, yshift=-1.5cm] (a1s3)
  {Generators\\(k4Gen)};

  \node[draw, rectangle, minimum height=2cm, minimum width=3cm, align=center, layer=green, xshift=5cm, yshift= 1.5cm] (a2s1) {Full Simulation\\\small ddsim, k4SimGeant4\\\small Gaussino};
  \node[draw, rectangle, minimum height=2cm, minimum width=3cm, align=center, layer=green, xshift=9cm, yshift= 1.5cm] (a2r1) {Reconstruction};
  \node[draw, rectangle, minimum height=2cm, minimum width=5cm, align=center, layer=green, xshift=7cm, yshift=-1.5cm] (a2s2) {Parametrized Simulation\\\small (k4Sim)Delphes};

  \node[draw, rectangle, minimum height=5cm, minimum width=3cm, align=center, layer=green, xshift=14cm, yshift=0cm] (a3s1) {Analysis};

  \draw[linew] (a1s1.east) .. controls (2.5,1.5) .. (a2s1.west) ;
  \draw[linew] (a1s2.east) .. controls (2.5,1.5) .. (a2s1.west) ;
  \draw[linew] (a1s3.east) .. controls (2.5,1.5) .. (a2s1.west) ;
  \draw[linew] (a1s1.east) .. controls (2.5,-1.5) .. (a2s2.west) ;
  \draw[linew] (a1s2.east) .. controls (2.5,-1.5) .. (a2s2.west) ;
  \draw[linew] (a1s3.east) .. controls (2.5,-1.5) .. (a2s2.west) ;

  \draw[linew] (a2s1.east) -- (a2r1.west);

  \draw[linew] (a2r1.east) .. controls (11.5,  1.5) .. (a3s1.west);
  \draw[linew] (a2s2.east) .. controls (11.5, -1.5) .. (a3s1.west);

  \draw[dashed] (a1.east) -- (a2.west);
  \draw[dashed] (a2.east) -- (a3.west);

\end{tikzpicture}
  \caption{Potential data flows via full or fast simulation.\label{fig:simflow}}
\end{figure}
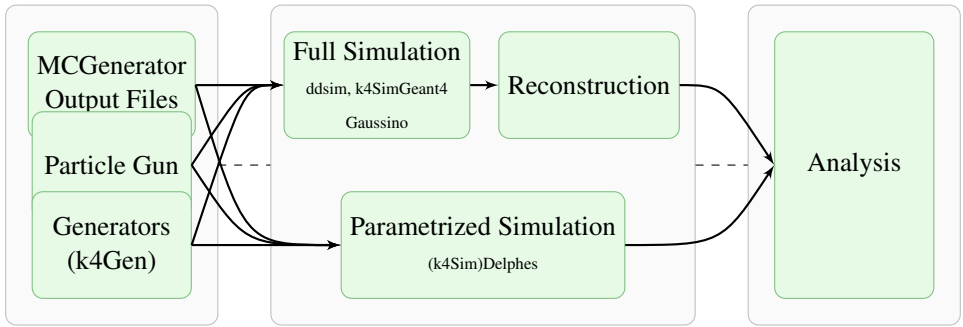

The Key4hep project provides a common framework for future Higgs factories and other experiments and has been fully
adopted by FCC and CLIC\@. It sees increasing adoption also from the ILC and CEPC communities. Beyond these initial
communities, the project has caught interest
of the EIC, Muon Collider communities, and LUXE experiment~\cite{LUXE_TDR}. To match the needs of the communities, the software
stack is expanding to state-of-the-art tools such as ACTS, PandoraPFA, CLUE,
or Phoenix. Their integrations, as outlined in the previous sections, that are or will be available soon as part of the
Key4hep stack, will allow its users to perform all the tasks needed for detector studies, as shown in \cref{fig:simflow}.

\section*{Acknowledgements}
This work beneﬁted from support by the CERN Strategic R\&D Programme on Technologies for Future Experiments
\linkk{https://cds.cern.ch/record/2649646/}{(CERN-OPEN-2018-006)}.
This project has received funding from the European Union's Horizon 2020 Research and Innovation programme under
Grant Agreement no. 101004761.
This project has received funding from the European Union's Horizon 2020 Research and Innovation programme under
Grant Agreement no. 871072.
This work has been sponsored by the Wolfgang Gentner Programme of the German Federal Ministry of Education and Research (grant no. 13E18CHA).

\bibliography{bibliography.bib}

\end{document}